\documentclass[%
 reprint,
 superscriptaddress,
 amsmath,amssymb,
 aps,
]{revtex4-1}

\usepackage{graphicx}% Include figure files
\usepackage{dcolumn}% Align table columns on decimal point
\usepackage{bm}% bold math
\usepackage{xcolor}
\begin{document}

%\preprint{APS/123-QED}

\title{The European Muon Collaboration effect from short-range correlated nucleons in a $x$-rescaling model}  %% Force line breaks with \\
%%\thanks{A footnote to the article title}%

\author{Rong Wang}
\email{rwang@impcas.ac.cn}
\affiliation{Institute of Modern Physics, Chinese Academy of Sciences, Lanzhou 730000, China}
\affiliation{School of Nuclear Science and Technology, University of Chinese Academy of Sciences, Beijing 100049, China}

\author{Na-Na Ma}
\email{mann@lzu.edu.cn}
\affiliation{School of Nuclear Science and Technology, Lanzhou University, Lanzhou 730000, China}

\author{Tao-Feng Wang}
\email{tfwang@buaa.edu.cn}
\affiliation{School of Physics, Beihang University, Beijing 100191, China}


\date{\today}% It is always \today, today,

\begin{abstract}
In this paper, we examine the hypothesis that the nuclear EMC effect comes
merely from the N-N SRC pairs inside the nucleus
and that the properties of N-N SRC pair are universal among the various nuclei,
using the conventional $x$-rescaling model for the EMC effect.
With the previously determined effective mass of the short-range correlated nucleon
and the number of N-N SRC pairs estimated,
we calculate the EMC effect of various nuclei within the $x$-rescaling approach.
From our calculations, the nuclear EMC effect due to the mass deficits
of the SRC nucleons is not enough to reproduce the observed EMC effect in experiments.
We speculate that the internal structure of the mean-field single nucleon is
also obviously modified, or there are more origins of the EMC effect
beyond the N-N SRC configuration (such as the $\alpha$ cluster),
or the universality of N-N SRC pair is violated noticeably from light to heavy nuclei.
\end{abstract}

%% \keywords{Suggested keywords}%Use showkeys class option if keyword
\maketitle

%\tableofcontents

\section{Introduction}
\label{sec:intro}

The European Muon Collaboration (EMC) effect,
one kind of nuclear medium modifications in the valence quark regime
of $0.3\lesssim x \lesssim 0.75$,
refers to the noticeable deviation from unity of the structure-function ratio
between the heavy nucleus ($A>2$) and the deuteron
\cite{EuropeanMuon:1983wih,EuropeanMuon:1988lbf,Bodek:1983qn,Bodek:1983ec,
Arneodo:1992wf,Geesaman:1995yd,Norton:2003cb,Hen:2013oha,Malace:2014uea}.
The deuteron here is taken as the reference nucleus,
which is approximately regarded as a system of a free proton plus a free neutron.
The EMC effect was firstly discovered in the muon-induced deep inelastic scattering
(DIS) off the iron nucleus \cite{EuropeanMuon:1983wih,EuropeanMuon:1988lbf}.
Soon after the discovery, the EMC effect was confirmed by the electron-nucleus DIS data
at SLAC \cite{Bodek:1983qn,Bodek:1983ec}.
Up to date, there are plenty of experimental measurements of the EMC effect on various nuclear targets.

The EMC effect is surprising and attracts a lot of interests from theorists.
The nuclear structure function is measured with the hard probe of momentum above GeV,
while the per-nucleon binding energy inside the nucleus is around the MeV scale.
At the fundamental level of quarks and gluons, we are not clear about
how the relatively long-range nuclear force modifies the short-distance structure inside the nucleon.
In the quark-parton model, the structure function is the incoherent
summation of the quark distributions inside the nucleon.
The discovery of the EMC effect implies that the quark distribution is
evidently modified by the nuclear medium.
If the quarks are completely confined inside the nucleon and do not play a role
in the emergence of nuclear force, then the quark distribution
should not be modified with the presence of surrounding nucleons.
With decades of studies, there are so many models constructed
which describe well the EMC effect, such as the off-shell correction \cite{Kulagin:1994fz,Kulagin:2004ie},
the $x$-rescaling model \cite{GarciaCanal:1984eh,Staszel:1983qx},
the nucleon swelling and dynamical rescaling model \cite{DiasdeDeus:1984wy,Wang:2016mzo,Wang:2018wfz,Close:1983tn,Jaffe:1983zw,Close:1984zn},
the cluster model \cite{Pirner:1980eu,Jaffe:1982rr,Carlson:1983fs,Date:1982na,Date:1984ve,DiasdeDeus:1984ge,Clark:1985qu,Barshay:1989ds},
the point-like configuration suppression model \cite{Frankfurt:1985cv,Frank:1995pv}, and the statistical model \cite{Zhang:2009vj}.
To differentiate the various models, more experiments and new observables
beyond the $F_2$ ratio are looking forward.

It is speculated that the strength of the EMC effect depends on
the local density instead of the global average density of the nucleus,
deduced from the measurements of the very light nuclei $^3$He and $^9$Be \cite{Seely:2009gt}.
This interesting finding stimulates the physicists to imagine that
the nuclear EMC effect is mainly from the local cluster structures inside the nucleus.
The other guess is that the EMC effect dominantly comes from the high-virtuality nucleons.
The high-virtuality nucleons belong to the short-distance configurations of nucleons.
This guess is supported by the amazing linear correlation between the EMC effect
and the nucleon-nucleon short-range correlations (N-N SRC) \cite{Weinstein:2010rt,Hen:2012fm}.

The N-N SRC pairs are the temporary close-proximity fluctuations
of two strongly interacting nucleons \cite{Hen:2016kwk,Arrington:2011xs,Frankfurt:1988nt,Fomin:2017ydn}.
In experiment, the N-N SRC are identified as the nucleon pair
of high relative momentum between nucleons and small center-of-mass momentum of the pair
\cite{Tang:2002ww,Subedi:2008zz,CLAS:2018yvt,CLAS:2018qpc,Arrington:2022sov}.
The nucleons in SRC have the momenta much higher than the nuclear Fermi momentum $k_{\rm F}$.
The abundance of N-N SRC pairs can be simply characterized by
the probability of finding the high-momentum nucleons \cite{CLAS:2005ola,Fomin:2011ng,CLAS:2019vsb}.
The nucleons of high momenta have the possibility to form the close-proximity configuration
and they are also sensitive to the repulsive core of nucleon-nucleon interaction \cite{CLAS:2020mom}.
It is widely accepted that the intermediate-distance tensor force
is the primary source of the formation of N-N SRC \cite{CLAS:2020mom,Schiavilla:2006xx,Alvioli:2007zz,Neff:2015xda}.

Inspired by the observed linear correlation between the EMC effect
and the N-N SRC \cite{Weinstein:2010rt,Hen:2012fm}, some nuclear physicists suggest that SRC pairs
may be the underlying source of the EMC effect.
This assumption actually is quite close to the traditional cluster model for the EMC effect
\cite{Pirner:1980eu,Jaffe:1982rr,Carlson:1983fs,Date:1982na,Date:1984ve,DiasdeDeus:1984ge,Clark:1985qu,Barshay:1989ds}.
The difference is that the cluster model is at the parton level with the six-quark bag picture,
and the SRC explanation is based on the nucleon degrees of freedom,
of which the properties of SRC nucleon are greatly modified.
In a recent theoretical work, people argue that the linear correlation
between the EMC effect and the N-N SRC is the natural result of
the scale separation of the nucleon structure part ($\Lambda$-independent)
and the twist-four part (nuclear modification, $\Lambda$-dependent) of nuclear matrix element \cite{Chen:2016bde}.
It is demonstrated that the linear correlation between the EMC effect and the N-N SRC
can be derived in the effective field theory.

One key and intriguing question is whether there is the causality between the EMC effect and SRC.
Recently, the CLAS collaboration tested the SRC-driven EMC model
with the simultaneous measurements of DIS and quasi-elastic inclusive process
on the deuteron and the heavier nuclei \cite{CLAS:2019vsb}.
They extracted the modification function of the nucleon structure in SRC pairs
and found that this modification function is nucleus-independent \cite{CLAS:2019vsb}.
They show that the EMC effect in all measured nuclei is consistent
with being due to the universal modification function of SRC pairs,
and that the magnitude of the EMC effect in the nucleus
can be described by the number of SRC pairs.
In their view, the EMC effect is not the traditional static modification
on all the independent nucleons but a strong dynamical effect for short time intervals
of two strongly interacting nucleons fluctuating into a temporary
high-local-density SRC pair \cite{CLAS:2019vsb}.
The universal modification function of SRC is also carefully studied
by J. Arrington and N. Fomin \cite{Arrington:2019wky}.
They found that there is almost no $A$-dependence
of the universal modification function extracted with Local-Density model,
while with the High-Virtuality model there is weak $A$-dependence.
The universal modification function from the data
of various nuclei is consistent with a truly universal function
and the Local-Density hypothesis is favored \cite{Arrington:2019wky}.

A different opinion is investigated and provided as well.
Recently, in Ref. \cite{Wang:2020uhj} people examine the relationship between SRC and the EMC effect
in more details by incorporating the nuclear binding and the nucleon off-shell effects.
They argue that their analysis does not support the hypothesis that
there is a causal connection between nucleons residing in SRCs and the EMC effect \cite{Wang:2020uhj}.
The EMC effect of the low-momentum nucleon and the high-momentum nucleon are studied separately.
They find that the Fermi motion effect overwhelms the off-shell effect for the SRC nucleons,
with three different models for the off-shell effect \cite{Wang:2020uhj}.
Hence they conclude that the SRC nucleons do not give the dominant EMC effect \cite{Wang:2020uhj}.
This conclusion is contrary to what one expected in the past.
It is worthwhile to further examine the relationship between the EMC effect
and SRC from different views or theoretical models.

Currently the approaches for describing the EMC effect can be classified into
the following three categories: (i) All the nucleons are slightly modified
when embedded in the nuclear medium; (ii) Nucleons are unmodified for most of
the time, but greatly modified when they fluctuate into N-N SRC; (iii)
Mean-field uncorrelated nucleons are slightly modified
and the nucleons are substantially modified for a short-time interval
in the temporary SRC state. It is a hot topic whether the nuclear EMC effect
entirely comes from the N-N SRC. Hence, in this work we focus on the
last two approaches in explaining the EMC effect. In Sec. \ref{sec:x-rescaling},
we illustrate the models we use to calculate the nuclear EMC effect.
In Sec. \ref{sec:results}, we present the results of the EMC effect from
SRC nucleons and mean-field nucleons. Finally, a short summary is given in Sec. \ref{sec:summary}.

\section{Nuclear EMC effect and $x$-rescaling model}
\label{sec:x-rescaling}

It is known that the traditional nuclear structure is almost
irrelevant to the nuclear EMC effect. The nucleon momentum distribution
leads to the Fermi motion effect around $x=1$.
The per-nucleon nuclear binding energy is much smaller than
the high-momentum virtual photon probe or the nucleon mass.
Nevertheless the identity of nucleons inside nucleus is well established
and it is the core of the traditional nuclear physics.
The nucleon structure should be connected to the properties of the nucleon.

In this work, we apply the $x$-rescaling model to evaluate the EMC effect,
which is based on the view that the nucleus is a sum of quasi-particles (bound nucleons).
The mass is a fundamental property of the nucleon,
and the change in the nucleon mass inside the nucleus
should be taken into account for the nuclear medium effect.
The nucleon effective mass in nucleus has been successfully
used to describe the nuclear EMC effect \cite{GarciaCanal:1984eh,Staszel:1983qx}.
The Bjorken scaling variable is defined in terms
of the free nucleon mass $m$, as $x=Q^2/(2m\nu)$.
However the struck nucleon in lepton-nucleus DIS could be far off-shell.
The true scaling variable for nuclear DIS should be taken to be
$x^{\prime}=Q^2/(2m^{*}\nu)=xm/m^{*}=x\eta$,
where $m^{*}$ is the effective mass of the bound nucleon.
Here $\eta=m/m^{*}$ is the rescaling factor of $x$.
Hence the per-nucleon nuclear structure function $F_2^{\rm A}$ is given by,
\begin{equation}
\begin{split}
F_2^{\rm A}(x, Q^2) = F_2^{\rm N}(x\eta, Q^2),
\end{split}
\label{eq:scaling-model}
\end{equation}
in which $F_2^{\rm N}$ is the free nucleon structure function.
The rescaling of $x$ is taken into account for the off-shell correction
of the bound nucleon \cite{GarciaCanal:1984eh,Staszel:1983qx,Akulinichev:1985ij}.
It was also pointed out that the exchanged virtual meson would 
take away a fraction of the nucleon momentum, thus resulting in
the $x$-rescaling of the nuclear structure function \cite{LlewellynSmith:1983vzz}.

The nucleon effective mass in the $x$-scaling model is used to describe the
off-shellness of the nucleon, with $E^2=p^2+m^{*2}$ \cite{GarciaCanal:1984eh,Staszel:1983qx,Akulinichev:1985ij}.
The nucleon effective mass was also defined by Brueckner in 1950s
in a non-relativistic many-body theory to account for the momentum-dependence
of potential energy of single particle, with $E(k)=k^2/2m + V(0)+bk^2=k^2/2m^{*}+V(0)$, $V(k)=V(0)+bk^2+...$,
and $m^{*}=m/(1+2bm)$ \cite{Brueckner:1955zze}. Therefore, the effective mass of SRC nucleon
in this work is different from Brueckner's definition.
Brueckner's nucleon effective mass reflects leading effects
of the space-time non-locality of the underlying nuclear interactions \cite{Li:2018lpy,Jeukenne:1976uy},
while the effective mass of SRC nucleon arises more from
the local interactions at short distance.
The relations between these two effective masses should be
investigated in the future.

\subsection{Model-A}
\label{subsec:model-A}

The intriguing question we try to answer in this paper is whether
the N-N SRCs are wholly responsible for the nuclear EMC effect.
Therefore we bring about the first model, referred as model-A for the convenience
of discussions, in which only the short-range correlated nucleons are substantially
modified while the uncorrelated nucleons are nearly unmodified.
This model strongly relies on the causality between SRC and the EMC effect,
i.e., the N-N SRC is the primary source of the EMC effect.
For model-A, the nuclear structure function $F_2^{\rm A}$ is decomposed as,
\begin{equation}
\begin{split}
F_2^{\rm A}=&\left[n^{\rm A}_{\rm SRC}F_2^{\rm p~in~SRC}+n^{\rm A}_{\rm SRC}F_2^{\rm n~in~SRC}\right. \\
&\left.+(Z-n^{\rm A}_{\rm SRC})F_2^{\rm p}+(A-Z-n^{\rm A}_{\rm SRC})F_2^{\rm n} \right] /A,
\end{split}
\label{eq:NuclearF2-modelA}
\end{equation}
where $n^{\rm A}_{\rm SRC}$ is the number of proton-neutron SRC pairs in nucleus $A$,
$F_2^{\rm p~in~SRC}$ and $F_2^{\rm n~in~SRC}$ are the modified nucleon structure functions
in the SRC pair, $F_2^{\rm p}$ and $F_2^{\rm n}$ are free nucleon structure functions.
In the formula, $Z$, $N$ and $A$ are respectively the proton number,
neutron number and the mass number.
Here the number of SRC pairs should be viewed as
the time-averaged value for the dynamical system.
Since the deuteron is just being in the SRC configuration occasionally,
the time-averaged number of SRC pairs in deuteron is less than one,
as $n^{\rm d}_{\rm SRC}<1$.

The SRC universality and the isophobic property of N-N SRC pair are
the other two foundations of model-A.
The universality of SRC can be described by the similar form of
nuclear wave function at high nucleon momentum,
which is confirmed by the experimental observations of the $x$-independence
and the weak $Q^2$-dependence of the cross section ratio
between two different nuclei in the region of $1.4\lesssim x \lesssim 2$ \cite{CLAS:2005ola,Fomin:2011ng,CLAS:2019vsb}.
Different experiments have revealed that most of the SRC pairs are the proton-neutron pairs
\cite{Subedi:2008zz,CLAS:2018yvt,Piasetzky:2006ai,Hen:2014nza,Fomin:2017ydn,Arrington:2022sov}.
This isophobic property supports the point that the immediate tensor force is
the primary source for the formation of N-N SRC pairs \cite{CLAS:2020mom,Schiavilla:2006xx,Alvioli:2007zz,Neff:2015xda}.

For model-A, the number of SRC pairs in nucleus $A$ and the
modified nucleon structure functions in SRC pair are the key inputs.
The number of SRC pairs in nucleus $A$ is closely related to the measured SRC scaling ratio $a_2$
(nucleus $A$ over deuteron) and the number of SRC pairs in deuteron, which is written as,
\begin{equation}
n^{\rm A}_{\rm SRC}=[A\times a_2(A)\times n_{\rm SRC}^{\rm d}]/2.
\label{eq:SRCPairNumber}
\end{equation}
Note that the above relation (Eq. (\ref{eq:SRCPairNumber})) is a simplified assumption.
The SRC scaling ratio $a_2$ are measured with the high-energy electron inclusive
scattering process off the nuclear targets \cite{CLAS:2005ola,Fomin:2011ng,CLAS:2019vsb}
and the number of SRC pairs in deuteron has already been determined in the previous analysis \cite{Wang:2020egq}.
The free nucleon structure functions can be calculated with the parton distribution functions $f_i(x,Q^2)$,
as $F_2^{\rm N}(x,Q^2) = \sum_i e_i^{2} xf_i(x,Q^2)$.
In this work, the proton parton distribution functions are taken from
the global analyses such as CT14 \cite{Dulat:2015mca} and CJ15 \cite{Accardi:2016qay}.
The parton distributions of the free neutron are easily given by the parton distributions
of the proton under the assumption of isospin symmetry, $u^{n}=d^{p}$ and $d^{n}=u^{p}$.
By using the $x$-rescaling model, the structure function of the SRC nucleon is connected
with the free nucleon structure function, which is written as,
\begin{equation}
\begin{split}
F_2^{\rm p~in~SRC}(x, Q^2) = F_2^{\rm p}(x\eta_{\rm SRC}, Q^2),\\
F_2^{\rm n~in~SRC}(x, Q^2) = F_2^{\rm n}(x\eta_{\rm SRC}, Q^2),\\
\end{split}
\label{eq:SRCNucleonF2}
\end{equation}
in which $\eta_{\rm SRC}$ is the rescaling factor for the SRC nucleon.
$\eta_{\rm SRC}$ is directly connected with the effective mass of SRC nucleon
as $\eta_{\rm SRC}=m/m_{\rm SRC}$, which is an universal factor among different nuclei.
Since the effective mass of SRC nucleon $m_{\rm SRC}$ has already been extracted
from a correlation analysis between the nuclear mass and SRC scaling ratio $a_2$,
the rescaling factor for SRC nucleon is computed to be $\eta_{\rm SRC}=1.10$ \cite{Wang:2020egq}.

\subsection{Model-B}
\label{subsec:model-B}

According to the nuclear shell model, the nucleons move independently
in the mutual potential created by all the nucleons, which is usually
approximated with the mean field. These mean-field nucleons are mainly
governed by the long-range nuclear force.
As revealed by the high energy electron probe, we know that the nucleon-nucleon
short-range correlations exist and they are the important microscopic structure of the nucleus.
Although the short-range correlated nucleons interact intensively,
they are the minorities being in a temporary state.
A more general hypothesis is that the structure function of mean-field nucleon
is slightly modified and the structure function of SRC nucleon is strongly modified.
In other words, the N-N SRC may not generate the enough EMC effect.

For the second model, referred as model-B for the convenience of discussions,
we propose that both the mean-field nucleons and the SRC nucleons are modified
by the nuclear medium or the correlated partner nucleon.
The nuclear structure function in model-B is decomposed as,
\begin{equation}
\begin{split}
F_2^{\rm A}=&\left[n^{\rm A}_{\rm SRC}F_2^{\rm p~in~SRC}+n^{\rm A}_{\rm SRC}F_2^{\rm n~in~SRC}\right. \\
&\left.+(Z-n^{\rm A}_{\rm SRC})F_2^{\rm p^*}+(A-Z-n^{\rm A}_{\rm SRC})F_2^{\rm n^*} \right] /A,
\end{split}
\label{eq:nuclearF2-modelB}
\end{equation}
where $F_2^{\rm p~in~SRC}$ and $F_2^{\rm n~in~SRC}$ denote the structure functions of SRC nucleons,
$F_2^{\rm p^*}$ and $F_2^{\rm n^*}$ denote the structure functions of mean-field nucleons.
Here, the number of SRC pairs $n^{\rm A}_{\rm SRC}$ and the structure functions of SRC nucleons
are taken as the same of the model-A.
In model-B, the structure functions of mean-field nucleons are also calculated with the $x$-rescaling model,
which is written as,
\begin{equation}
\begin{split}
F_2^{\rm p^*}(x, Q^2) = F_2^{\rm p}(x\eta_{\rm MF}, Q^2),\\
F_2^{\rm n^*}(x, Q^2) = F_2^{\rm n}(x\eta_{\rm MF}, Q^2).\\
\end{split}
\label{eq:MFNucleonF2}
\end{equation}
Different from the situation for SRC nucleon, we assume that the rescaling factor $\eta_{\rm MF}$
for mean-field nucleon is nucleus-dependent, since the effective mass of mean-field nucleon
depends on the nucleus. The nucleon densities of different nuclei are different.
In this analysis, we let $\eta_{\rm MF}$ be a free parameter for each nucleus.
Note that the rescaling factor $\eta_{\rm MF}$ for mean-field nucleon should be smaller
than the rescaling factor $\eta_{\rm SRC}$ for SRC nucleon.

\section{Results and discussions}
\label{sec:results}

\begin{figure}[htp]
\centering
\includegraphics[width=0.42\textwidth]{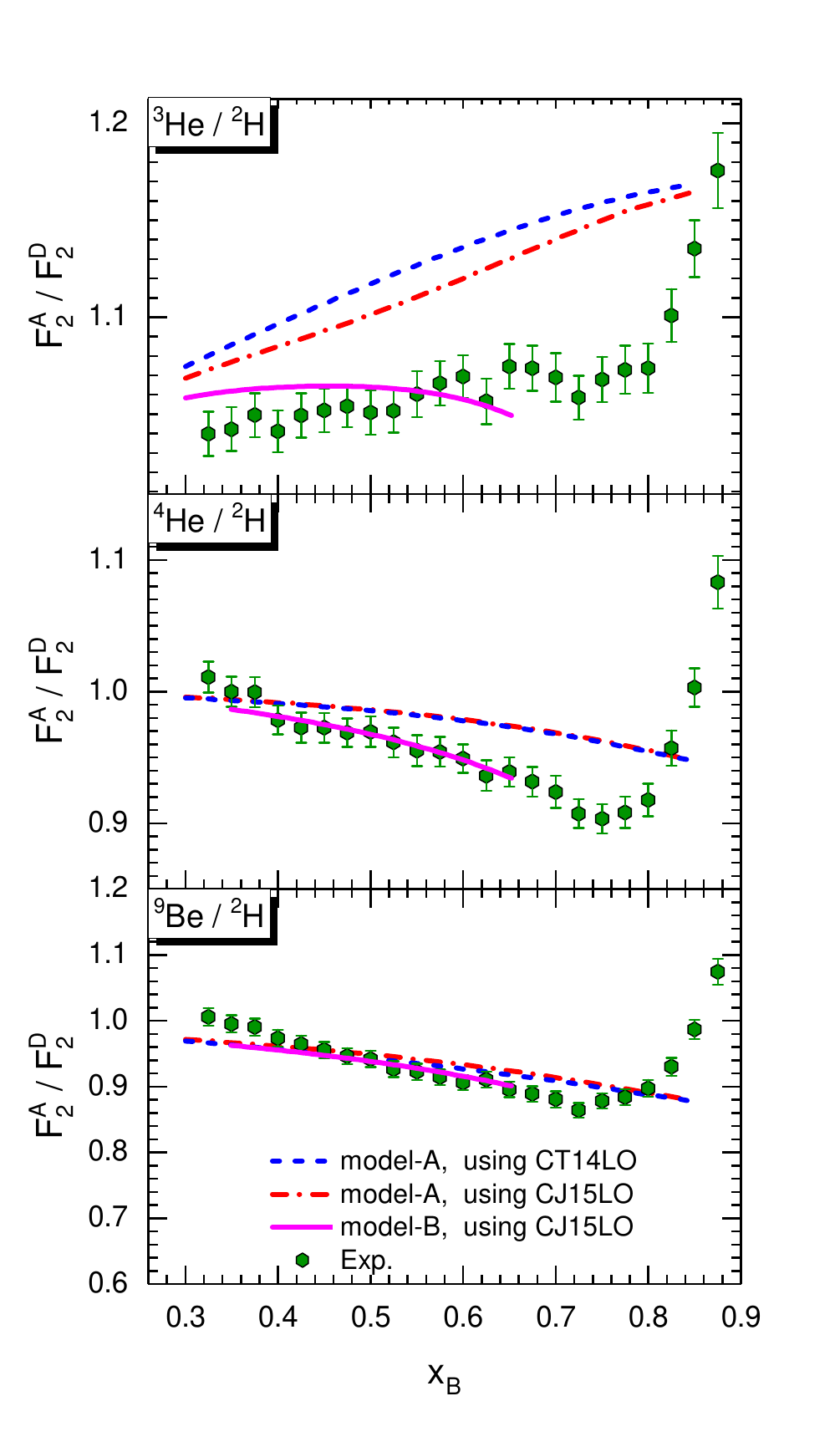}
\caption{
The predicted EMC ratios from the $x$-rescaling models are shown with the experimental data (light nuclei).
See the main text for the details of the models.
The experimental data are taken from JLab Hall C \cite{Seely:2009gt}.
$Q^2$ is 5.3 GeV$^2$ in the model calculations, to be consistent with the experiment.
}
\label{fig:EMCLightNuclei}
\end{figure}

\begin{figure*}[htp]
\centering
\includegraphics[width=0.75\textwidth]{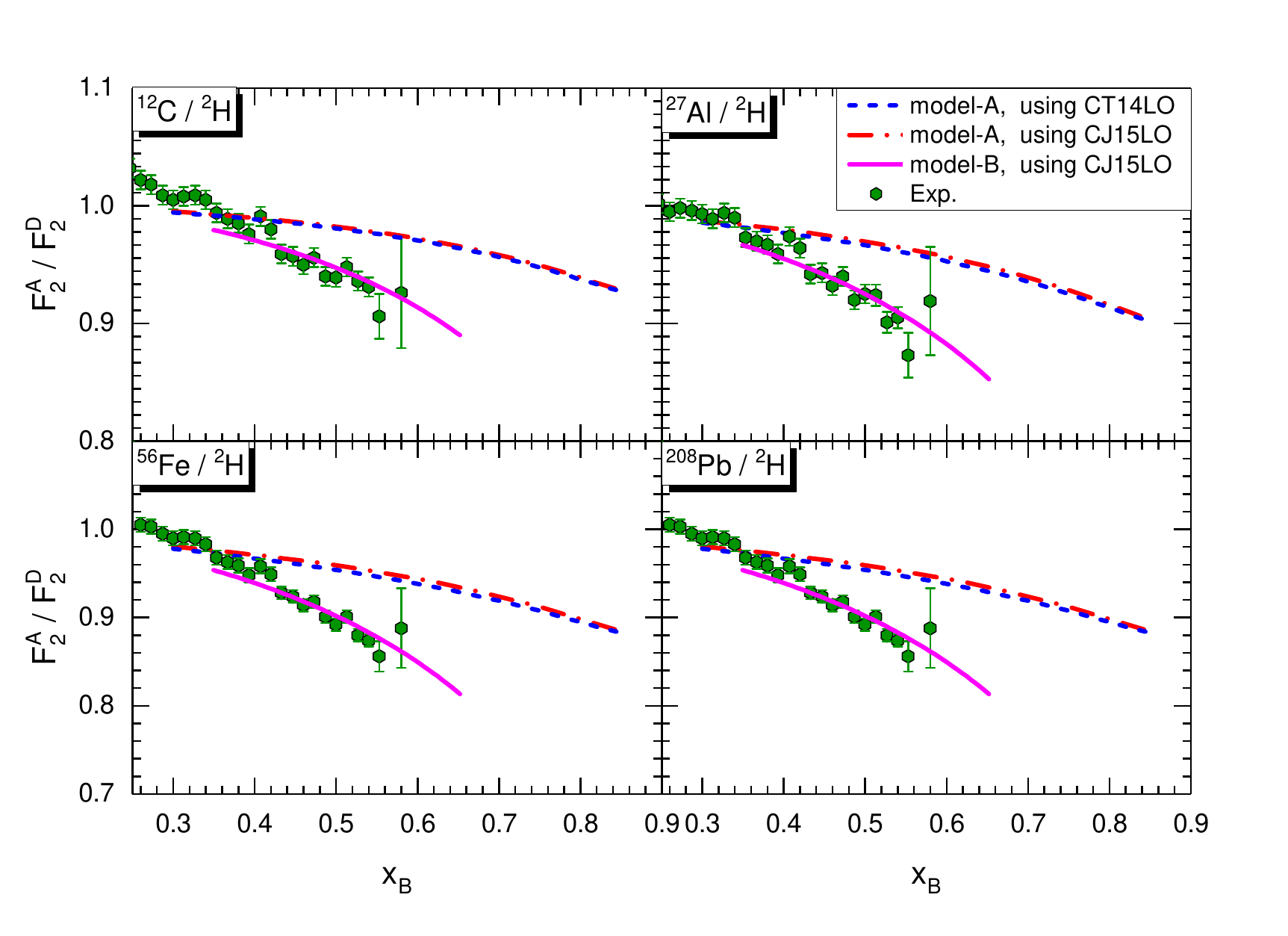}
\caption{
The predicted EMC ratios from the $x$-rescaling models are shown with the experimental data (heavy nuclei).
See the main text for the details of the models.
The experimental data are taken from CLAS at JLab \cite{CLAS:2019vsb}.
$Q^2$ is 2 GeV$^2$ in the model calculations, to be consistent with the experiment.
}
\label{fig:EMCHeavyNuclei}
\end{figure*}

\begin{figure*}[htp]
\centering
\includegraphics[width=0.75\textwidth]{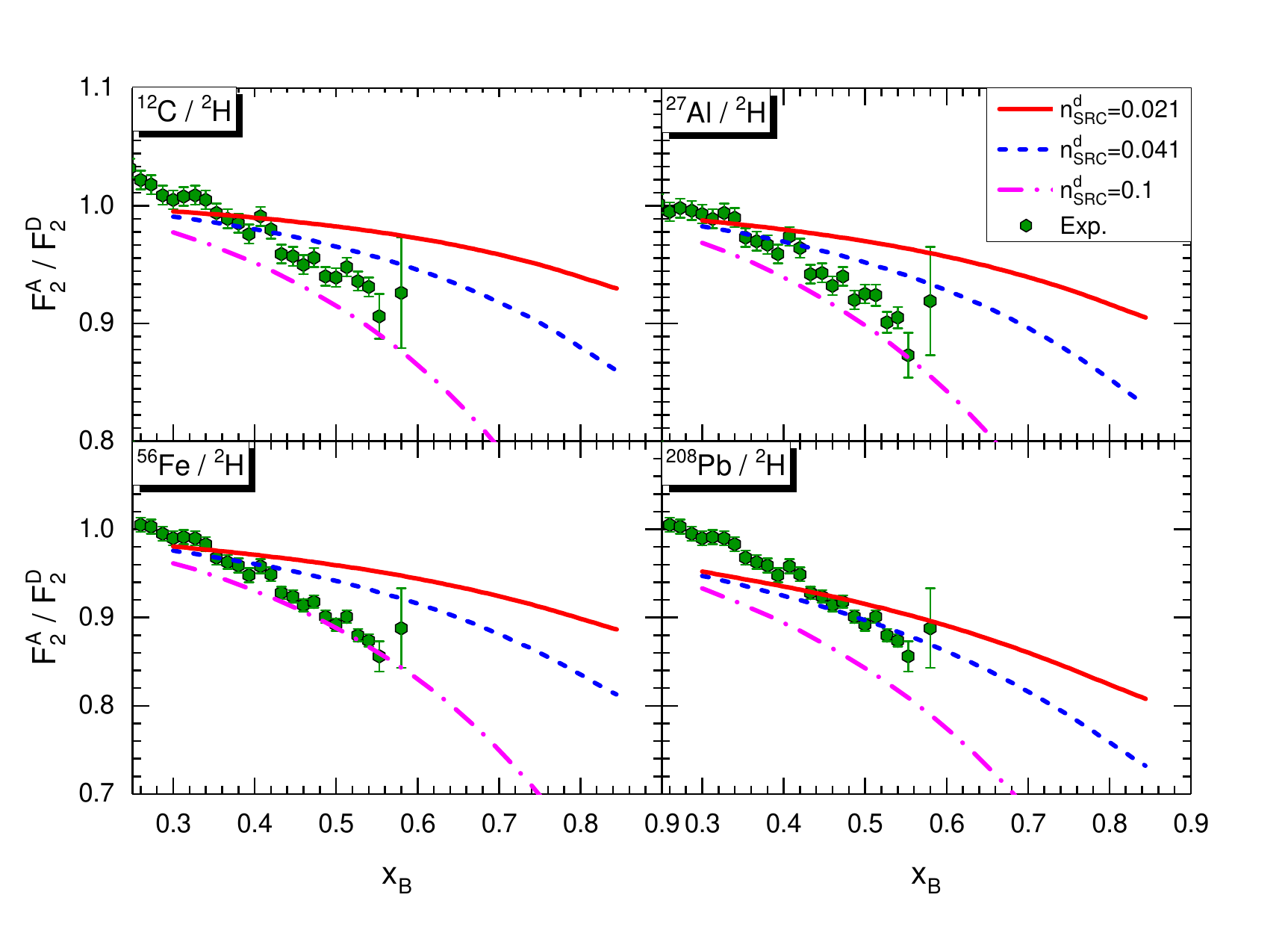}
\caption{
The predicted EMC ratios from a $x$-rescaling model (model-A) are shown
with the experimental data (heavy nuclei), with different input values
for the parameter $n^{\rm d}_{\rm SRC}$.
See the main text for the details of the model.
The experimental data are taken from CLAS at JLab \cite{CLAS:2019vsb}.
$Q^2$ is 2 GeV$^2$ in the model calculations, to be consistent with the experiment.
}
\label{fig:EMCHeavyNuclei-more-SRCs}
\end{figure*}

\begin{table}[h]
\caption{
The fitted rescaling factor $\eta_{MF}$ for the mean-field nucleon are listed,
under the framework of model-B.
In the model, modifications on both the SRC nucleon and the mean-field nucleon
consequence in the observed nuclear EMC effect.
The errors come only from the fits to the EMC effect data.
The uncertainties of the parameters $n^{\text{d}}_{\text{SRC}}$ and $a_2$ are not included.
}
\begin{center}
\begin{tabular}{ cc|cc }
  \hline\hline
  ~~~nucleus~~~&~~~~~~$\eta_{\text{MF}}$~~~~~~ &~~~nucleus~~~& ~~~~~~$\eta_{\text{MF}}$~~~~~~       \\
  \hline
  $^4$He        &  1.008 $\pm$ 0.001     & $^9$Be        &  1.005 $\pm$ 0.002    \\
  $^{12}$C      &  1.016 $\pm$ 0.002     & $^{27}$Al     &  1.021 $\pm$ 0.002    \\
  $^{56}$Fe     &  1.027 $\pm$ 0.001     & $^{208}$Pb    &  1.022 $\pm$ 0.002    \\
  \hline\hline
\end{tabular}
\end{center}
\label{tab:rescaling-factor-MF}
\end{table}

\begin{figure}[htp]
\centering
\includegraphics[width=0.42\textwidth]{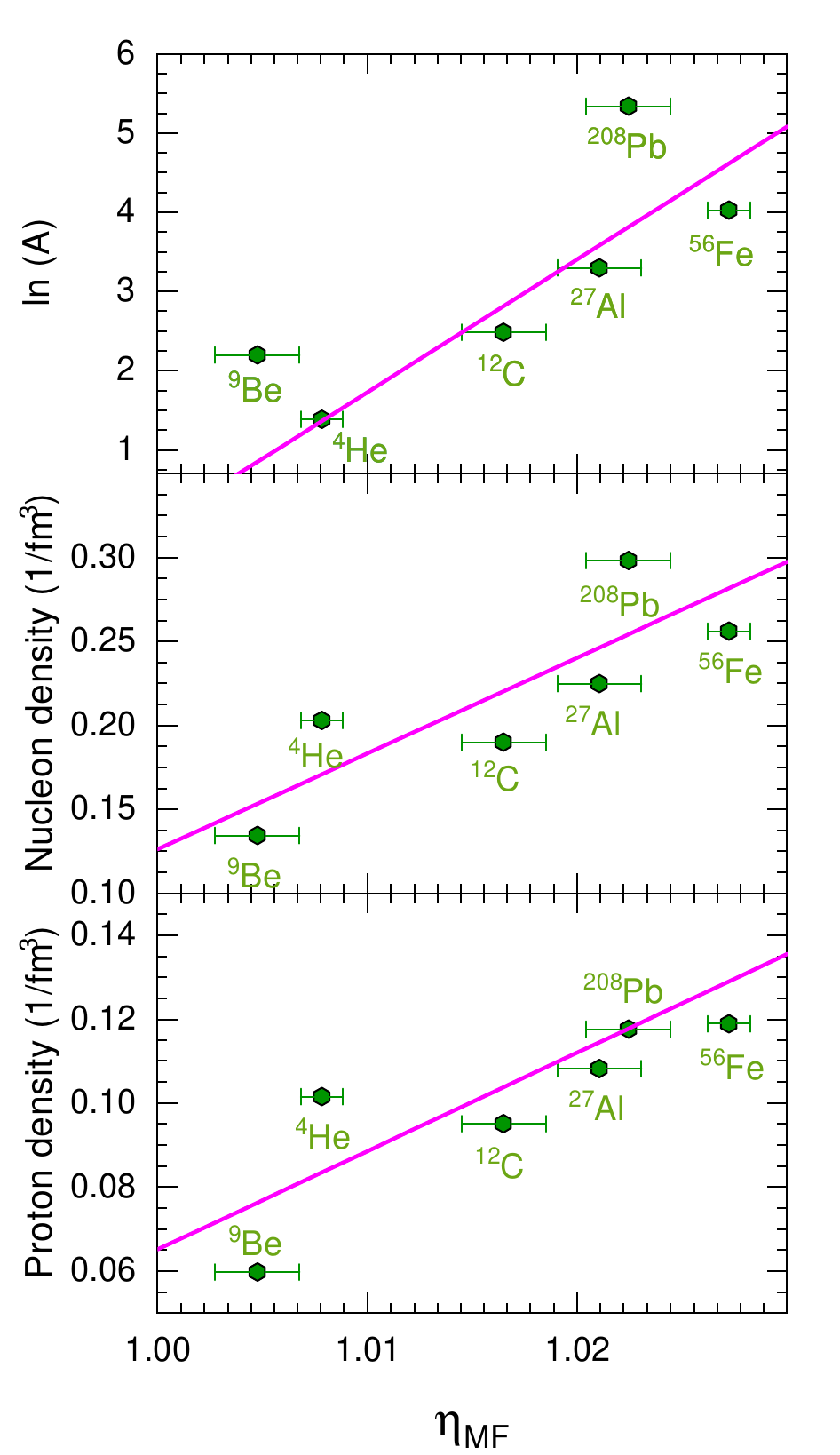}
\caption{
The correlations among the rescaling factor $\eta_{MF}$ of mean-field nucleon,
$ln(A)$, the average nucleon density, and the average proton density.
See the main text for how these densities are calculated.
}
\label{fig:etaMeanField}
\end{figure}

Fig. \ref{fig:EMCLightNuclei} and \ref{fig:EMCHeavyNuclei} depict
the recent experimental measurements on the nuclear EMC effects
in light nuclei and heavy nuclei, respectively.
The predictions of model-A and model-B are also shown in the figures for comparisons.
The experimental data are taken from the recent high-precision measurements by CLAS at JLab \cite{CLAS:2019vsb}.
We see that the experimental data points are distributed
in the valence quark region of $x$ smaller than 0.6.
As the data is quite away from the Fermi motion region near $x\sim 1$,
the Fermi motion correction is neglected in this work.

We find that the EMC effect from model-A is much weaker than the experimental observations.
For the calculations of structure function ratios in model-A,
the parton distribution functions of CT14 and CJ15 are used.
We see that the dependence on the data set of parton distribution functions is weak.
In conclusion, based on the $x$-rescaling model,
the nuclear modifications from short-range correlated nucleons only
are not enough to reproduce the nuclear EMC effect in experiments.
We speculate that either the valence distribution of mean-field nucleon is also modified,
or the modification of valence distribution in N-N SRC is not universal in different nuclei,
or some other short-distance structures beyond N-N SRC exist with strong modifications
on the inner nucleon structure, such as 3N-SRC and $\alpha$ clusters.

We notice that the number of proton-neutron SRC pairs in deuteron is estimated
to be $n^{\rm d}_{\rm SRC}=0.041$ by K. S. Egiyan et al. \cite{CLAS:2005ola},
which is much larger than the value from our previous analysis \cite{Wang:2020egq}.
In their analysis, the number of nucleons in N-N SRC pairs is defined as
the number of nucleons of high momenta $k>k_{\rm F}\approx 275$ MeV/c \cite{Subedi:2008zz,CLAS:2005ola}.
With this definition, a small fraction of mean-field nucleons may be misidentified
as SRC nucleons, resulting in more SRC pairs than our previous analysis.
Nevertheless, in Fig. \ref{fig:EMCHeavyNuclei-more-SRCs}, we show the predicted EMC ratios from model-A,
taking the SRC $a_2$ data averaged from experiments \cite{CLAS:2019vsb,Hen:2012fm}
and $n^{\rm d}_{\rm SRC}=0.041$ \cite{CLAS:2005ola}.
The predicted EMC slopes are still smaller than the data by CLAS collaboration.
Therefore, based on either our estimation on SRC numbers or the estimation by K. S. Egiyan et al.,
the modifications on SRC nucleons only are not enough to interpret the nuclear EMC effect.
Then we let the number of SRC pairs inside deuteron be a free parameter.
And we find that as the number of SRC pairs inside deuteron increases to as many as 10\%,
the EMC effect can be explained with only the SRC nucleons.
However, such many SRC pairs inside deuteron is contradictory (much higher than)
to the analysis based on the experimental data.

In model-B, the mean-field nucleons are also modified, in addition to the SRC nucleons.
We also assume in model-B that the rescaling factor $\eta_{\rm MF}$
is a free parameter and it depends on the nuclear medium.
Thus we performed the least square fits of model-B
to the EMC ratio data in the range of $0.35<x_B<0.65$,
to find the optimal parameter $\eta_{\rm MF}$ for each measured nucleus.
The obtained $\eta_{\rm MF}$ are listed in Table \ref{tab:rescaling-factor-MF}.
In model-B, $\eta_{MF}$ of deuteron is simply one.
The $\eta_{MF}$ of $^{208}$Pb is determined to be 1.022 $\pm$ 0.002,
and it is much smaller than the rescaling factor for the SRC nucleon.
Nevertheless, the mean-field nucleons in $^{208}$Pb are evidently modified,
judged by the obtained rescaling factor $\eta_{MF}$.
By introducing the EMC effect of the mean-field nucleon,
the model-B successfully explains the nuclear EMC effect.

Furthermore, let us take a look at the nuclear dependence of
the rescaling factor $\eta_{MF}$ for the mean-field nucleon in model-B.
The correlations between $\eta_{MF}$ and $ln(A)$, $\eta_{MF}$ and the nucleon density,
$\eta_{MF}$ and the proton density are shown in Fig. \ref{fig:etaMeanField}.
The nucleon density and proton density are calculated using
$A/(\frac{4}{3}\pi R^3)$ and $Z/(\frac{4}{3}\pi R^3)$ respectively,
in which $R$ is the charge radius of a nucleus.
The data of nuclear charge radii are taken from Ref. \cite{Angeli:2013epw}.
Since the radius of the neutron distribution in the nucleus may not be the same as the charge radius,
that is why we also plot the correlation between $\eta_{MF}$ and the proton density of the nucleus.
Although the linear correlation is not perfect, the rescaling factor $\eta_{MF}$
of the mean-field nucleon is more or less correlated with the nucleon density.
The obtained rescaling factor of the mean-field nucleon is proportional to the average nuclear density.

\section{Summary and outlook}
\label{sec:summary}

In the $x$-rescaling model, we have tested the idea that
the N-N SRC is the dominant source for the nuclear EMC effect.
The nuclear EMC effects of some nuclei are calculated within the $x$-re scaling model,
under the assumptions that the SRC nucleon is universal among different nuclei
and that only the inner structure of short-range correlated nucleons are modified.
The input mass of the N-N SRC pair and the number of SRC pairs inside the deuteron
are taken from the previous analysis of the $a_2$ data and nuclear mass \cite{Wang:2020egq}.
We find that the nuclear medium correction on N-N SRC is not enough to explain
the EMC effect observed in experiments, if the model applied in this work is correct.
This conclusion is consistent with a result analyzed with the off-shellness correction \cite{Wang:2020uhj}.

If we assume that the rescaling factor $\eta_{\rm SRC}$ is $A$-dependent,
then Model-A could describe well the experimental data.
However this assumption breaks the universality of N-N SRC
that is basically supported by experimental observations \cite{CLAS:2005ola,Fomin:2011ng,CLAS:2019vsb}
and some theoretical predictions \cite{Feldmeier:2011qy,Alvioli:2016wwp}.
For Model-B, we find that the rescaling factor for uncorrelated
nucleon is roughly linearly correlated with the nuclear density.
This linear relation can be tested with further experimental
measurements on more nuclear targets of different densities.

We speculate that more origins of nucleon structure modifications beyond
the short-distance configurations are needed, such as 3N SRC and $\alpha$ clusters.
Another possible interpretations are either the mean-field nucleon is noticeably modified,
or the N-N SRC pairs in different nuclei have different nuclear medium modifications.
For a preliminary exploration, We have shown that the EMC effect can be explained
if we just assume that the mean-field nucleon is also modified.
And the nuclear modification on the mean-field nucleon scales with the density of the nucleus.
In summary, we conclude that either the SRC universality is wrong,
or the mean-field nucleon is also modified slightly,
or there are other sources beyond N-N SRC for the EMC effect, such as the $\alpha$ cluster,
or the applied $x$-rescaling model needs improvement.

The strong evidence of 3N SRC is not found in the inclusive $^4$He/$^3$He cross section
ratio at JLab, and it is shown that isolating 3N SRC is much more challenging compared
to 2N SRC \cite{HallA:2017ivm}. However, the theorists suggest that the scaling phenomenon from
inclusive scattering on 3N SRC requires a high $Q^2\gtrsim 3$ GeV$^2$
and the current experimental situation should be improved \cite{Day:2018nja}.
Within Model-A, $n^{\rm d}_{\rm SRC}$ needs to increase from 0.021 to 0.1,
in order to explain the experimental data of the EMC effect.
Considering that the 3N SRC and the $\alpha$ cluster also contribute to the EMC effect,
the numbers of 3N SRC pairs and $\alpha$ clusters should be at the same order of N-N SRC
if the structure-function modifications of N-N SRC, 3N SRC and $\alpha$ are similar.
Since there is no 3N SRC and $\alpha$ cluster in deuteron
and that the nuclear modifications inside 3N SRC and $\alpha$ could be stronger
than that inside N-N SRC, the numbers of 3N SRC pairs and $\alpha$ clusters in heavy nuclei
could be smaller than that of N-N SRC pairs, inferred from the current data of the EMC effect.
More experimental measurements are needed for searching other short-distance structures
beyond N-N SRC.

\begin{acknowledgments}
This work is supported by the National Natural Science Foundation of China under the Grant NO. 12005266
and the Strategic Priority Research Program of Chinese Academy of Sciences under the Grant NO. XDB34030301.
N.-N. Ma is supported by the National Natural Science Foundation of China under the Grant NO. 12105128.
T.-F. Wang is supported by the National Natural Science Foundation of China under the Grant NOS. 10175091 and 11305007.
\end{acknowledgments}

\bibliographystyle{apsrev4-1}
\bibliography{refs}

\end{document}